\documentclass{article}
\pdfoutput=1
\usepackage[utf8]{inputenc}
\usepackage{amsmath}
\usepackage{hyperref}
\usepackage{titling}
\usepackage[english]{babel}
\usepackage{underscore}
\usepackage{fullpage}
\usepackage{listings}
\usepackage{parskip}
\usepackage{authblk}
\usepackage{float}
\usepackage[title]{appendix}
\usepackage{etoolbox}
\usepackage{changelog}

\patchcmd{\appendices}{\quad}{: }{}{}

\newcommand{\subtitle}[1]{%
  \posttitle{%
    \par\end{center}
    \begin{center}\large#1\end{center}
    \vskip0.5em}%
}

\newcommand{\email}[1]{\href{mailto:#1}{\nolinkurl{#1}}}

\title{Third DIHARD Challenge Evaluation Plan}
\subtitle{Version 1.2}
\date{October 30, 2020}
\author[a]{Neville Ryant}
\author[b]{Kenneth Church}
\author[a]{Christopher Cieri}
\author[c]{Jun Du}
\author[d]{Sriram Ganapathy}
\author[a]{Mark Liberman}
\affil[a]{Linguistic Data Consortium, University of Pennsylvania, Philadelphia, PA, USA}
\affil[b]{Baidu Research, Sunnyvale, CA, USA}
\affil[c]{University of Science and Technology of China, Hefei, China}
\affil[d]{Electrical Engineering Department, Indian Institute of Science, Bangalore, India}

\begin{document}
\maketitle

\begin{changelog}[simple,sectioncmd=\section*]
\begin{version}[v=1.2,
    date=10-30-2020]
    \item Planned schedule updated
    \item Updated Table~\ref{tab:single_chan_dev_set} and Table~\ref{tab:single_chan_eval_set} with numbers from final versions of the development/evaluation sets
    \item Added link to results submission instructions on DIHARD III website
    \item Corrected typo in Section~\ref{sec:intro} regarding the amount of new clinical data
\end{version}
\begin{version}[v=1.1,
    date=07-13-2020]
    \item Planned schedule updated
    \item Minor updates to registration instructions
    \item Updated workshop venue and date
    \item Added additional training resources to appendix \ref{app:data}
\end{version}
\end{changelog}

\section{Introduction}
\label{sec:intro}
DIHARD III is the third in a series of diarization challenges focusing on ``hard'' diarization; that is, speaker diarization for challenging recordings where there is an expectation that the current state-of-the-art will fare poorly. As with other evaluations in this series, DIHARD III is intended to both (1) support speaker diarization research through the creation and distribution of novel data sets and (2) measure and calibrate the performance of systems on these data sets. The results of the challenge will be presented at online workshop to be held January 23rd, 2021.

The task evaluated in the challenge is speaker diarization; that is, the task of determining ``who spoke when'' in a multispeaker environment based only on audio recordings. As with DIHARD I and II, development and evaluation sets will be provided by the organizers, but there is no fixed training set with the result that participants are free to train their systems on any proprietary and/or public data. Once again, these development and evaluation sets will be drawn from a diverse sampling of sources including monologues, map task dialogues, broadcast interviews, sociolinguistic interviews, meeting speech, speech in restaurants, clinical recordings, and YouTube videos. However, there are several key differences from DIHARD II:
    \begin{itemize}
        \item the diarization from multi-channel audio condition that was evaluated as part of DIHARD II will not be evaluated this year; parties interested in this condition should instead consult the results from track 2 of CHiME-6\footnote{\url{https://chimechallenge.github.io/chime6/}}, which is essentially  a rerun of the DIHARD II multichannel condition
        \item the child speech domain present in DIHARD I and II has been removed; while this is an extremely interesting domain, issues with the license restrictions on the data and the nature of the task (diarization of child speech is almost a different task from diarization of adult speech) necessitated its removal for this year's evaluation
        \item 4 hours of previously unexposed audio has been added to the clinical domain
        \item for the first time, we are including conversational telephone speech  --  20 hours of previously unexposed 2 person English language calls selected from the unreleased Fisher English Phase 2 collection
        \item for the first time, DIHARD is partnering with NIST through the OpenSAT evaluation series\footnote{\url{https://www.nist.gov/itl/iad/mig/opensat}}; all evaluation activities (registration, system submission, scoring, and leaderboard display) will be conducted using NIST maintained web-interfaces
        \item instead of an Interspeech special session, results of the challenge will be presented at a post-evaluation workshop; due to continued COVID-19 impacts, this workshop will be held online
    \end{itemize}

Participation in the evaluation is open to all who are interested and willing to comply with the rules laid out in this evaluation plan. There is no cost to participate and the web interface, data, scoring software, and any baselines are provided free of charge. Participating teams will have the option of attending the post-evaluation workshop\footnote{\url{https://dihardchallenge.github.io/dihard3workshop/}}, to be held online on January 23rd, 2021. Information about evaluation registration can be found on the DIHARD III website\footnote{\url{https://dihardchallenge.github.io/dihard3}}.

For questions not answered in this document or to join the DIHARD mailing list, please contact \email{dihardchallenge@gmail.com}.

\section{Planned Schedule}
\label{sec:schedule}
	\begin{itemize}
        \item registration opens  -- September 10th, 2020
    	\item development set release  --  September 21st, 2020
    	\item evaluation set release  --  September 21st, 2020
    	\item scoring server opens  --  October 30th, 2020
    	\item evaluation period ends  --  December 21st, 2020 (midnight Anywhere on Earth)
    	\item abstract submission deadline  --  December 21st, 2020 (midnight Anywhere on Earth)
    	\item system descriptions due  --  January 8th, 2020 (midnight Anywhere on Earth)
    	\item workshop  --  January 23rd, 2021
    \end{itemize}

\section{Task}
\subsection{Task definition}
The goal of the challenge is to automatically detect and label all speaker segments in each recording session. Small pauses of $<=$ 200 ms by a speaker are not considered to be segmentation breaks and should be bridged into a single continuous segment. A pause by a speaker is defined as any segment in which that speaker is not producing a vocalization of any kind. By vocalization, we mean speech, including speech errors and infant babbling, but also vocal noise such as breaths, coughs, lipsmacks, sneezes, laughs, humming or any other noise produced by the speaker by means of the vocal apparatus.

\subsection{Tracks}
Because  system  performance  is  strongly  influenced  by  the  quality  of  the speech segmentation used, two different tracks are covered:
    \begin{itemize}
        \item {\bf Track 1}  --  Diarization from reference SAD. Systems are provided with a reference speech segmentation that is generated by merging speaker turns in the reference diarization.
        \item {\bf Track 2}  --  Diarization from scratch. Systems are provided with just the raw audio input for each recording session and are responsible for producing their own speech segmentation.
    \end{itemize}

\subsection{Evaluation conditions}
\label{sec:task:test}
For DIHARD III, we define two partitions of the evaluation data:
    \begin{itemize}
        \item {\bf core evaluation set}  --  a ``balanced'' evaluation set in which the total duration of each domain\footnote{See Section \ref{sec:data} and appendix \ref{app:data} for more details regarding the domains.} is approximately equal
        \item {\bf full evaluation set}  --  a larger evaluation set that uses all available selections for each domain and which is, thus, unbalanced with some domains having more audio than others; it is a proper superset of the core evaluation set
    \end{itemize}
The core evaluation set strives for balance across domains so that the evaluation metrics are not dominated by any single domain. It mimics the evaluation set composition from DIHARD I and II. The full evaluation set includes additional material from two domains (CLINICAL and CTS), potentially resulting in more stable metrics at the expense of being unbalanced. All system submissions to all tracks will be scored against both sets and the results reported on the leaderboards.

\section{Scoring}
System output will be scored by comparison to human reference segmentation with performance evaluated by two metrics:
    \begin{itemize}
        \item diarization error rate (DER)
        \item Jaccard error rate (JER) 
    \end{itemize}

\subsection{Diarization error rate}
Diarization error rate (DER), introduced for the NIST Rich Transcription Spring 2003 Evaluation (RT-03S)\footnote{\url{https://web.archive.org/web/20160805233512/http://www.itl.nist.gov/iad/mig/tests/rt/2003-spring/index.html}}, is the total percentage of reference speaker time that is not correctly attributed to a speaker, where ``correctly attributed'' is defined in terms of an optimal mapping between the reference and system speakers. More concretely, DER is defined as: 
\begin{equation*}
    \textrm{DER} = \frac{\textrm{FA} + \textrm{MISS} + \textrm{ERROR}}{\textrm{TOTAL}}
\end{equation*}
where
\begin{itemize}
    \item {\it TOTAL} is the total reference speaker time; that is, the sum of the durations of all reference speaker segments
    \item {\it FA} is the total system speaker time not attributed to a reference speaker
    \item {\it MISS} is the total reference speaker time not attributed to a system speaker
    \item {\it ERROR} is the total reference speaker time attributed to the wrong speaker
\end{itemize}
Contrary to practice in the NIST RT evaluations, {\bf NO} forgiveness collar will be applied to the reference segments prior to scoring and overlapping speech {\bf WILL} be evaluated. For more details please consult section 6 of the RT-09 evaluation plan\footnote{\url{https://web.archive.org/web/20100606041157if_/http://www.itl.nist.gov/iad/mig/tests/rt/2009/docs/rt09-meeting-eval-plan-v2.pdf}} and the source to the NIST {\it md-eval} scoring tool\footnote{Available as part of the Speech Recognition Scoring Toolkit (SCTK): \url{ftp://jaguar.ncsl.nist.gov/pub/sctk-2.4.10-20151007-1312Z.tar.bz2}. For DIHARD, we will be using version 22 of {\it md-eval}.}.

\subsection{Jaccard error rate}
In addition to the primary metric we will score systems using Jaccard error rate (JER), first introduced in DIHARD II. The Jaccard error rate is based on the Jaccard index\footnote{\url{https://en.wikipedia.org/wiki/Jaccard_index}}, a similarity measure used to evaluate the output of image segmentation systems. An optimal mapping between reference and system speakers is determined and for each pair the Jaccard index is computed. The Jaccard error rate is then defined as 1 minus the average of these scores. While similar to DER, it weights every speaker's contribution equally, regardless of how much speech they actually produced.

More concretely, assume we have $N$ reference speakers and $M$ system speakers. An optimal mapping between speakers is determined using the Hungarian algorithm so that each reference speaker is paired with at most one system speaker and each system speaker with at most one reference speaker. Then, for each reference speaker $ref$ the speaker-specific Jaccard error rate $JER_{ref}$ is computed as:
    \begin{equation*}
        \textrm{JER}_{ref} = \frac{\textrm{FA} + \textrm{MISS}}{\textrm{TOTAL}}
    \end{equation*}
where
\begin{itemize}
    \item {\it TOTAL} is the duration of the union of reference and system speaker segments; if the reference speaker was not paired with a system speaker, it is the duration of all reference speaker segments
    \item {\it FA} is the total system speaker time not attributed to the reference speaker; if the reference speaker was not paired with a system speaker, it is 0
    \item {\it MISS} is the total reference speaker time not attributed to the system speaker; if the reference speaker was not paired with a system speaker, it is equal to {\it TOTAL}
\end{itemize}
The Jaccard error rate then is the average of the speaker specific Jaccard error rates:
    \begin{equation*}
        \textrm{JER} = \frac{1}{N}\sum_{ref}\textrm{JER}_{ref}
    \end{equation*}
As with DER {\bf NO} forgiveness collar will be applied to the reference segments prior to scoring and overlapping speech {\bf WILL} be evaluated.

JER and DER are highly correlated with JER typically being higher, especially in recordings where one or more speakers is particularly dominant. Where it tends to track DER is in outliers where the diarization is especially bad, resulting in one or more unmapped system speakers whose speech is not then penalized. In these cases, where DER can easily exceed 500\%, JER will never exceed 100\% and may be far lower if the reference speakers are handled correctly.

\subsection{Scoring regions}
In most cases the scoring region for each recording will be the {\bf entirety} of the recording; that is, for a recording of duration 405.37 seconds, the scoring region will be [0, 405.37]. However, for a small subset of the recordings, personal identifying information (PII) has been removed from the recording, either by low-pass filtering or insertion of tones or zeroing out of samples. For these recordings, the scoring regions consist of the entirety of the recording minus these regions. In both cases the scoring regions will be specified by un-partitioned evaluation map (UEM) files, which will be distributed by LDC as part of the development and evaluation releases. Please see Appendix \ref{app:uem} for details of the UEM file format.

\subsection{Scoring tool}
All scoring will be performed using version 1.0.1 of {\it dscore}, which is maintained as a github repo at:
    \begin{quote}
        \url{https://github.com/nryant/dscore}
    \end{quote}
To score a set of system output RTTMs {\it sys1.rttm}, {\it sys2.rttm}, ... against corresponding reference RTTMs {\it ref1.rttm}, {\it ref2.rttm}, ... using the un-partitioned evaluation map (UEM) {\it all.uem}, the command line would be:
\begin{lstlisting}[language=bash]
  $ python score.py -u all.uem -r ref1.rttm ref2.rttm ... -s sys1.rttm sys2.rttm ...
\end{lstlisting}
The overall and per-file results for DER and JER (and many other metrics) will be printed to STDOUT as a table. For additional details about scoring tool usage, please consult the documentation for the github repo.

\section{Data}
\label{sec:data}
\subsection{Training data}
\label{sec:data:train}
Participants may use any publicly available and/or proprietary data to train their systems, with the exception of the following previously released corpora, from which portions of the evaluation set are drawn:
\begin{itemize}
    \item DCIEM Map Task Corpus (LDC96S38)
    \item MIXER6 Speech (LDC2013S03)
    \item Digital Archive of Southern Speech (LDC2012S03 and LDC2016S05)
    \item DIHARD I and II evaluation sets
\end{itemize}
Portions of MIXER6 have previously been excerpted for use in the NIST SRE10\footnote{\url{https://www.nist.gov/system/files/documents/itl/iad/mig/NIST_SRE10_evalplan-r6.pdf}} and SRE12\footnote{\url{https://www.nist.gov/system/files/documents/itl/iad/mig/NIST_SRE12_evalplan-v17-r1.pdf}} evaluation sets, which also may not be used.

All training data should be thoroughly documented in the system description document (see Appendix \ref{app:system}) at the end of the challenge. For a list of suggested training corpora, please consult Appendix \ref{app:data}.

\subsection{Development and evaluation data}
\label{sec:singlechannel}
The development and evaluation sets consist of  selections of 5-10 minute duration samples\footnote{Excepting data drawn from the WEB VIDEO domain, which range from under 1 minute to more than 10 minutes.} drawn from 11 domains. For most domains, the same source is used for both the development and evaluation sets, though in some cases the development and evaluation sets use different sources; where the two sets draw from different sources, this is noted. For a detailed explanation of the domains and sources, please consult Appendix \ref{app:sources}.

\subsubsection{Development data}
A development set is provided that mirrors the composition of the evaluation set and which may be used for any purpose, including system training. The full composition of this development set, including domains, the sources drawn on for each domain, and total duration for the core set and full set\footnote{The distinction between ``core set'' and ``full set'' is identical to the test set, as discussed in Section \ref{sec:task:test}.} is presented in Table \ref{tab:single_chan_dev_set}.
\begin{table}[H]
    \centering
        \begin{tabular}{llrr}
        \hline
         {\bf Domain}               & {\bf Source} &  {\bf Core set (hours)} &   {\bf Full set (hours)} \\
        \hline
         AUDIOBOOKS                 & LIBRIVOX     &  2.01                   &        2.01 \\
         BROADCAST INTERVIEW        & YOUTHPOINT   &  2.06                   &        2.06 \\
         CLINICAL                   & ADOS         &  2.06                   &        4.27 \\
         COURTROOM                  & SCOTUS       &  2.08                   &        2.08 \\
         CTS                        & FISHER       &  2.17                   &       10.17 \\
         MAP TASK                   & DCIEM        &  2.53                   &        2.53 \\
         MEETING                    & RT04         &  2.45                   &        2.45 \\
        RESTAURANT                  & CIR          &  2.03                   &        2.03 \\
         SOCIOLINGUISTIC (FIELD)    & SLX          &  2.01                   &        2.01 \\
         SOCIOLINGUISTIC (LAB)      & MIXER6       &  2.67                   &        2.67 \\
         WEB VIDEO                  & VAST         &  1.89                   &        1.89 \\
         \hline
         TOTAL                      & -            &  23.94                  &       34.15 \\
        \hline
        \end{tabular}
    \caption{Composition of the core and full development sets. For explanation of domains and sources, consult Appendix \ref{app:sources}.}
    \label{tab:single_chan_dev_set}
\end{table}
For each recording, the following metadata is provided:
    \begin{itemize}
        \item the domain
        \item the source drawn from
        \item the language
        \item whether or not it was selected for the core development set
    \end{itemize}

\subsubsection{Evaluation data}
The full composition of the evaluation sets, including domains, the sources drawn on for each domain, and total duration for the core set and full set is presented in Table \ref{tab:single_chan_eval_set}. Note that this set uses different sources than the development set for two domains:
    \begin{itemize}
        \item the MEETING domain draws from ROAR instead of RT04
        \item the SOCIOLINGUISTIC (FIELD) domain draws from DASS instead of SLX
    \end{itemize}
The following metadata will {\bf NOT} be provided for the recordings during the evaluation period, but will be revealed at the conclusion of the evaluation:
    \begin{itemize}
        \item the domain
        \item the source drawn from
        \item the language
        \item whether or not it was selected for the core evaluation set
    \end{itemize}

\begin{table}[H]
    \centering
        \begin{tabular}{llrr}
        \hline
         {\bf Domain}                & {\bf Source}  &  {\bf Core set (hours)}   &   {\bf Full set (hours)} \\
        \hline
         AUDIOBOOKS                  & LIBRIVOX      &  2.04                     &    2.04 \\
         BROADCAST INTERVIEW         & YOUTHPOINT    &  2.03                     &    2.03\\
         CLINICAL                    & ADOS          &  2.08                     &    4.36 \\
         COURTROOM                   & SCOTUS        &  2.04                     &    2.04 \\
         CTS                         & FISHER        &  2.17                     &   10.17 \\
         MAP TASK                    & DCIEM         &  2.07                     &    2.07 \\
         MEETING                     & ROAR          &  1.87                     &    1.87 \\
        RESTAURANT                   & CIR           &  2.06                     &    2.06 \\
         SOCIOLINGUISTIC (FIELD)     & DASS          &  2.27                     &    2.27 \\
         SOCIOLINGUISTIC (LAB)       & MIXER6        &  2.03                     &    2.03\\
         WEB VIDEO                   & VAST          &  2.07                     &    2.07 \\
         \hline
         TOTAL                       & -             & 22.73                     &   33.01 \\
        \hline
        \end{tabular}
    \caption{Composition of the full and core evaluation sets. For explanation of domains and sources, consult Appendix \ref{app:sources}.}
    \label{tab:single_chan_eval_set}
\end{table}

\subsubsection{Segmentation}
\label{sec:single_track_segmentation}
Reference diarization was produced by segmenting the recordings into labeled speaker turns according to the following guidelines:
\begin{itemize}
    \item split on pauses $>$ 200 ms, where a pause by speaker ``S'' is defined as any segment of time during which ``S'' is not producing a vocalization of any kind, where vocalization is defined as any noise produced by the speaker by means of the vocal apparatus\footnote{For instance, speech (including yelled and whispered speech), backchannels, filled pauses, singing, speech errors and disfluencies, infant babbling or vocalizations, laughter, coughs, breaths, lipsmacks, and humming.}
    \item attempt to place boundaries within 10 ms of the true boundary, taking care not to truncate sounds at edges of words (e.g., utterance-final fricatives or utterance initial stops)
    \item where close-talking microphones exist for each speaker (e.g., ROAR), perform the segmentation separately for each speaker using their individual microphone
\end{itemize}
Reference SAD was then derived from these segmentations by merging overlapping speech segments and removing speaker identification.

During DIHARD II, it was found that manual annotation to this spec required use of highly skilled and experienced annotators using multiple spectrogram displays, making the annotation extremely slow and costly. Many annotators were incapable of performing the task even after extensive training and the remainder (usually people with experience in both ASR and acoustic phonetics) found it extremely laborious with real time rates typically greater than 15X and sometimes exceeding 30X\footnote{Recordings from VAST and SEEDLINGS (removed for DIHARD III) were found to be particularly difficult.}. Consequently, for DIHARD III we abandoned a commitment to entirely manual segmentation. Where a manual segmentation to these specs already exists (i.e., files annotated for DIHARD II), we use it. For all other data we instead produce a careful turn-level transcription (if one does not exist), then establish boundaries using a forced aligner trained using Kaldi. For DIHARD III, forced alignment was used for the following domains:
\begin{itemize}
    \item CTS
    \item CLINICAL
\end{itemize}
For all other domains we reused the manual annotation produced for DIHARD II.

\subsubsection{PII}
A limited number of recordings from ADOS, CIR, and DASS contained regions carrying  personal identifying information (PII), which had to be removed prior to publication. As systems have no way of plausibly dealing with these regions, they will not be scored and the relevant UEM files reflect this.  The method used to de-identify these regions differs from source to source, with some opting to replace PII containing regions with a pure tone, while others used an approach based on low-pass filtering. Please see Appendix \ref{app:sources} for details about how PII was dealt with for each source.

\subsubsection{File formats}
All audio and annotations will be distributed via LDC. The audio will be distributed as single channel, 16 bit FLAC files sampled at 16 kHz, while reference speech segmentations will be distributed as HTK label files. In the case of the development set, a reference diarization will be provided, which will be distributed as  Rich Transcription Time Marked (RTTM) files. For details regarding these file formats, please see Appendix \ref{app:sad} and Appendix \ref{app:rttm}.

\section{Evaluation rules}
\label{sec:rules}
There is no cost to participate in the DIHARD evaluation series. Participation is open to all who comply with the evaluation rules set forth in this plan. Development data and evaluation data will be made available to registered participants via LDC. Participants will apply their systems to the evaluation data locally and upload their system outputs to the DIHARD scoring server\footnote{Generously hosted by NIST's OpenSAT program.} for scoring. 

All participants agree to process the data in accordance with the following rules:
\begin{itemize}
    \item Participants agree to make at least one {\bf valid} primary system submission to track 1 before the end of the evaluation period. A valid submission is defined as one that contains RTTMs for all recordings and passes the validation step during upload.
    \item While most of the test data is actually, or effectively, unexposed, portions have been exposed in part in the following corpora:
    	\begin{itemize}
            \item DCIEM Map Task Corpus (LDC96S38)
            \item MIXER6 Speech (LDC2013S03)
            \item Digital Archive of Southern Speech (LDC2012S03 and LDC2016S05)
			\item NIST SRE10 evaluation data
            \item NIST SRE12 evaluation data
            \item DIHARD I and II evaluation sets
        \end{itemize}
        Participants agree to not use these corpora for system training or development.
    \item Manual/human investigation of the evaluation set (e.g., listening, segmentation, or transcription) prior to the end of the evaluation is disallowed.
    \item Participants are allowed to use any automatically derived information  (e.g., automatic identification of the domain) for the development and evaluation files provided that the systems used were not trained using any of the prohibited corpora.
    \item Participants may make multiple primary and contrastive submissions during the evaluation period (up to 50 valid submissions for each track\footnote{This limit is for primary and contrastive submissions {\bf COMBINED}. For instance, a team could make 15 primary and 35 contrastive submissions to track 1 or 40 primary and 10 contrastive submissions, but not 15 primary and 40 contrastive submissions.}).  For each track the leaderboards will display for each team the result of the the most recently processed valid primary system submission. 
\end{itemize}

In addition to the above data processing rules, the participants agree to comply with the following general requirements:
\begin{itemize}
    \item Participants agree to submit {\bf by the designated deadline} (see Section \ref{sec:schedule}) a system description document describing the algorithms, data, and computational resources used for systems. These documents will be submitted at the end of the evaluation and should follow the format set forth in Appendix \ref{app:system}.
    \item Participants agree to allow the deposit of the RTTM outputs of their final primary system outputs (i.e., those displayed on the leaderboards at the end of the evaluation) on Zenodo. At the conclusion of the challenge, the organizers will deposit an archive on Zenodo containing all system descriptions and final system outputs.
\end{itemize}
\vspace{0.25cm}
\emph{\textbf{Sites failing to abide by the above rules will be excluded from future evaluation participation and their registrations will not be accepted until they are committed to fully participate.}}

\section{Evaluation protocol}
All evaluation activities will be conducted over a NIST maintained web-interface to facilitate information exchange between evaluation participants and the organizers.

\subsection{Setting up an evaluation account}
Participants must sign up for an evaluation account, which will allow them to perform various activities such as registering for the evaluation, agreeing to the evaluation terms and conditions, signing the data license agreement, and uploading submissions. To sign up for an evaluation account, follow the instructions at: 
    \begin{quote}
        \url{https://dihardchallenge.github.io/dihard3/registration.html}
    \end{quote}
After the evaluation account is confirmed, the participant will be asked to join a site (or create one if it does not exist). The participant is also asked to associate their site to a team or to create a team if one does not exist. This allows multiple members to perform activities on behalf of their site and/or team (e.g., make a submission). Clarifying the distinction between participants, sites, and teams:
    \begin{itemize}
        \item site  --   a single organization (e.g., NIST)
        \item team  --  a group of organizations collaborating on a task (e.g., Team1 consisting of NIST and LDC)
        \item participant  --  a member or representative of a site who takes part in the evaluation (e.g., John Doe)
    \end{itemize}

\subsection{Evaluation registration}
One participant from a site must formally register their site to participate in the evaluation by agreeing to the terms of participation (see Section \ref{sec:rules}) and selecting the tasks they wish to participate in. 
For additional instructions, consult the \href{https://dihardchallenge.github.io/dihard3/registration.html}{DIHARD III website}.

\subsection{Data license agreement}
One participant from each {\bf site} must sign and upload the LDC data license agreement. After the license agreement is confirmed by LDC, LDC will provide instructions for accessing the development and evaluation data. For additional instructions, consult the \href{https://dihardchallenge.github.io/dihard3/registration.html}{DIHARD III website}.

\subsection{Results submission}
All system outputs must be submitted via the evaluation dashboard. Step-by-step instructions for submitting system outputs for scoring are available on  the \href{https://dihardchallenge.github.io/dihard3/submission.html}{DIHARD III website}.

\section{Workshop}
The results of the challenge will be presented on January 23rd, 2021 at an online workshop. Teams wishing to submit 2 page extended abstracts to this workshop should follow the instructions at:
    \begin{quote}
        \url{https://dihardchallenge.github.io/dihard3workshop/}
    \end{quote}

\section{Updates}
Updates to this evaluation plan will be made available via the mailing list and the challenge website (\url{https://dihardchallenge.github.io/dihard3}).

\newpage
\begin{appendices}
\section{Single Channel Condition Domains and Sources}
\label{app:sources}

\subsection*{Domains}
    \begin{itemize}
        \item {\it Audiobooks} \\
            Excerpts from recordings of speakers reading aloud passages from public domain English language texts. The recordings were selected from LibriVox and each recording consists of a single, amateur reader. Care was taken to make sure that the chapters and speakers drawn from were not present in LibriSpeech, which also draws from LibriVox. 
        \item {\it Broadcast interview} \\
            Student-lead radio interviews conducted during the 1970s with popular figures of the era (e.g., Ann Landers, Mark Hamill, Buckminster Fuller, and Isaac Asimov). The recordings are selected from the unpublished LDC YouthPoint corpus.
        \item {\it Clinical} \\
            Recordings of Autism Diagnostic Observation Schedule (ADOS) interviews conducted to identify whether a child fit the clinical diagnosis for autism. ADOS is a roughly hour long semi-structured interview in which clinicians attempt to elicit language that differentiates children with Autism Spectrum Disorder from those without (e.g., ``What does being a friend mean to you?''). The children included in this collection ranged from 12-16 years in age and exhibit a range of diagnoses from autism to non-autism language disorder to ADHD to typically developing. Interviews are typically recorded for quality assurance purposes; in this case, the recording was conducted using a ceiling mounted microphone. The recordings are selected from the unpublished LDC ADOS corpus.
        \item {\it Courtroom} \\
            Recordings of oral arguments from the 2001 term of the U.S. Supreme Court. The original recordings were made using individual table-mounted microphones, one for each participant, which could be switched on and off by the speakers as appropriate. The outputs of these microphones were summed and recorded on a single-channel reel-to-reel analogue tape recorder. All recordings are taken from SCOTUS, an unpublished LDC corpus.
        \item {\it CTS} \\
            Conversational telephone speech (CTS) consisting of 10 minute conversations between two native English speakers. All calls are drawn from the unreleased Phase II calls from the Fisher English collection conducted as part of the DARPA EARS project.
        \item {\it Map task} \\
            Recordings of pairs of speakers engaged in  a map task. Each map task session contains two speakers sitting opposite one another at a table. Each speaker has a map visible only to him and a designated role as either ``Leader'' or ``Follower''. The Leader has a route marked on his map and is tasked with communicating this route to the Follower so that he may precisely reproduce it on his own map. Though each speaker was recorded on a separate channel via a close-talking microphone, these have been mixed together for the DIHARD releases. The recordings are drawn from the DCIEM Map Task Corpus (LDC96S38).
        \item {\it Meeting} \\
            Recordings of meetings containing between 3 and 7 speakers. The speech in these meetings is highly interactive in nature consisting of large amounts of spontaneous speech containing frequent interruptions and overlapping speech. For each meeting a single, centrally located distant microphone is provided, which may exhibit excessively low gain. For the development set, these meetings are drawn from RT04, while for the evaluation set they are drawn from ROAR.
        \item {\it Restaurant} \\
            Informal conversations recorded in restaurants using binaural microphones. Each session contains between 4 and 7 speakers seated at the same table at a restaurant at lunchtime and was recorded from a binaural microphone worn by a designated facilitator; the mix of the two channels recorded by this microphone are provided. This data exhibits the following properties, which are expected to make it particularly challenging for automated segmentation and recognition:
                \begin{itemize}
                    \item due to the microphone setup, the majority of the speakers are farfield
                    \item background speech from neighboring tables is often present, sometimes at levels close to that of the primary speakers in the conversation
                    \item background noise is abundant with clinking silverware, moving chairs/tables, and loud music all common
                    \item the conversations are informal and highly interactive with interruptions and frequent overlapped speech
                \end{itemize}
            All data is taken from LDC's unpublished CIR corpus.
        \item {\it Sociolinguistic field recordings} \\
           Sociolinguistic interviews recorded under field conditions. Recordings consists of a single interviewer attempting to elicit vernacular speech from an informant during informal conversation. Typically, interviews were recorded in the home, though occasionally they were recorded in a public location such as a park or cafe. The development set recordings were drawn from SLX and the evaluation set from DASS.
        \item {\it Sociolinguistic lab recordings} \\
            Sociolinguistic interviews recorded under quiet conditions in a controlled environment. All data is taken from the PZM microphones of LDC's Mixer 6 collection (LDC23013S03).
        \item {\it Web video} \\
            English and Mandarin amateur videos collected from online video sharing sites  (e.g., YouTube and Vimeo). This domain is expected to be particularly challenging as the videos present a diverse set of topics and recording conditions; in particular,  many videos contain multiple speakers talking in a noisy environment, where it can be difficult to distinguish speech from other kinds of sounds. All data is selected from LDC's VAST collection.
    \end{itemize}

\subsection*{Sources}
    \begin{itemize}
        \item {\it ADOS} \\
            ADOS is an unpublished LDC corpus consisting of transcribed excerpts from ADOS interviews conducted at the Center for Autism Research (CAR) at the Children's Hospital of Philadelphia (CHOP). All interviews were conducted at CAR by trained clinicians using ADOS module 3. The interviews were recorded using a mixture of cameras and audio recorded from a ceiling mounted microphone. Portions of these interviews determined by a clinician to be particularly diagnostic were then segmented and transcribed.
            
            Note that in order to publish this data, it had to be de-identified by applying a low-pass filter to regions identified as containing personal identifying information (PII). Pitch information in these regions is still recoverable, but the amplitude levels have been reduced relative to the original signal. Filtering was done with a 10th order Butterworth filter with a passband of 0 to 400 Hz. To avoid abrupt transitions in the resulting waveform, the effect of the filter was gradually faded in and out at the beginning and end of the regions using a ramp of 40 ms.
        \item {\it CIR} \\
            Conversations in Restaurants (CIR) is a collection of informal speech recorded in restaurants that LDC originally produced for the NSF Hearables Challenge\footnote{\url{https://www.nasa.gov/feature/nsf-hearables-challenge}}, an NSF-sponsored challenge designed to promote the development of algorithms or methods that could improve hearing in a noisy setting. It consists of conversations between 3 and 6 speakers, all LDC or Penn employees, seated at the same table at a restaurant near the University of Pennsylvania campus. Recording sessions were held at lunch time using a rotating list of restaurants exhibiting diverse acoustic environments and typically lasted 60-70 minutes. All recordings were conducted using binaural microphones mounted on either side of one speaker's head. 
        
            A limited number of regions from one recording were found to contain PII. These regions were de-identified using the same low-pass filtering approach as in ADOS.
        \item {\it DASS} \\
            The Digital Archive of Southern Speech, or DASS, is a corpus of interviews (each lasting anywhere from 3 to 13 hours) recorded during the late 60s and 70s in the Gulf Coast region of the United States. It is part of the larger Linguistic Atlas of the Gulf States (LAGS), a long-running project that attempted to preserve the speech of a region encompassing Louisiana, Alabama, Mississippi, and Florida as well as parts of Texas, Tennessee, Arkansas, and Georgia. Each interview was conducted in the field by a trained interviewer, who attempted to elicit conversation about common topics like family, the weather, household articles, agriculture, and social connections. It is distributed by LDC as LDC2012S03 and LDC2016S05.
            
            Due to the nature of the interviews, they sometimes contain PII or sensitive materials. All such regions have been replaced by tones of matched duration. Unfortunately, this process does not appear to have been systematic, with the result that the type of tone (pure or complex), power, and frequency differs across the corpus.
        \item {\it DCIEM} \\
            The DCIEM Map Task Corpus (LDC96S38) is a collection of recordings of two-person map tasks recorded for the DCIEM Sleep Deprivation Study. This study was conducted by the Defense and Civil Institute of Environmental  Medicine  (Department  of National  Defense,  Canada) to evaluate the effect of drugs on performance degradation in sleep deprived individuals. Three drug conditions (Modafinil vs. Amphetamine vs. placebo) were crossed with three sleep conditions (18 hours vs. 48 hours vs. 58 hours awake). During each session, subjects performed a battery of neuropsychological tests (e.g., tracking tasks, time estimation tasks, attention-splitting tasks), questionnaires, and a map task. All audio was recorded via close-talking microphones under quiet conditions.
        \item {\it LibriVox} \\
            LibriVox\footnote{\url{https://librivox.org/}} is a collection of public domain audiobooks read by volunteers from around the world. It consists of more than 10,000 recordings in 96 languages. Portions have previously appeared in the popular LibriSpeech\footnote{\url{http://www.openslr.org/12/}} corpus, though care was taken to ensure that DIHARD did not select from this subset.
        \item {\it FISHER} \\
            The Fisher corpora\footnote{\url{https://www.ldc.upenn.edu/sites/www.ldc.upenn.edu/files/lrec2004-fisher-corpus.pdf}} are a series of conversational telephone speech (CTS) collections undertaken as part of the DARPA EARS (Effective, Affordable, Reusable Speech-to-text) program. Each session consisted of a conversation between two randomly assigned (potentially non-native) speakers, who were prompted to talk for up to ten minutes on a randomly assigned topic. A wide range of topic prompts were used, examples of which include:
                \begin{itemize}
                    \item Do either of you have a favorite TV sport? How many hours per week do you spend watching it and other sporting events on TV?
                    \item Do you think most people would remain calm, or panic during a terrorist attack? How do you think each of you would react?
                    \item Do you like cold weather or warm weather activities the best? Do you like outside or inside activities better? Each of you should talk about your favorite activities.
                \end{itemize}
            All calls were recorded by LDC and transcribed by a combination of LDC and BBN.

            The Fisher English collection was performed in two phases, the first of which, comprising approximately 2,000 hours of audio from 12,000 speakers, has been released by LDC under catalog entries LDC2004S13, LDC2004T19, LDC2005S13, and LDC2005T19. A further 1,400 hours (from 1,300 speakers) were collected and transcribed during Phase II, but never released. For DIHARD III, we draw from these previously unexposed recordings.
        \item {\it MIXER6} \\
            Mixer 6 (LDC2013S03) is a large-scale collection of English speech across multiple environments, modalities, degrees of formality, and channels that was conducted at LDC from 2009 through 2010. The collection consists of interviews with 594 native speakers of English spanning 1,425 sessions, each roughly 40-45 minutes in duration. Each session contained multiple components (e.g., informal conversation styled after a sociolinguistic interview or transcript reading) and was captured by a variety of microphones, including lavalier, head-mounted, podium, shotgun, PZM, and array microphones. While the corpus was released without speaker segmentation or transcripts, a portion of the corpus was subsequently transcribed at LDC. DIHARD II draws its selections from this subset.
        \item {\it ROAR} \\
            ROAR is a collection of multiparty (3 to 6 participant) conversations recorded by LDC as part of the DARPA ROAR (Robust Omnipresent Automatic Recognition) project in Fall 2001. While portions of this collection have previously been exposed during the NIST RT evaluations, all DIHARD data comes from previously unexposed meetings. The meetings were recorded at LDC in a purpose built room using a combination of lavalier, head mounted, omnidirectional, PZM, shotgun, podium, and array microphones. For each meeting, a single centrally located distant microphone is provided.
        \item {\it RT04} \\
            RT04 consists of meeting speech released as part of the NIST Spring 2004 Rich Transcription (RT-04S) Meeting Recognition Evaluation development and evaluation sets. This data was later re-released by LDC as LDC2007S11 and LDC2007S12. It consists of recordings of multiparty (3 to 7 participant) meetings held at multiple sites (ICSI, NIST, CMU, and LDC), each with a different microphone setup. For DIHARD, a single channel is distributed for each meeting, corresponding to the RT-04S single distant microphone (SDM) condition. Audio files have been trimmed from the original recordings to the 11 minute scoring regions specified in the RT-04S un-partitioned evaluation map (UEM) files\footnote{In cases where the onset or offset of a scoring region was found to bisect a speaker turn, it was adjusted to fall in silence adjacent to the relevant turn.}.
        \item {\it SCOTUS} \\
            SCOTUS is an unpublished LDC corpus consisting of oral arguments from the 2001 term of the U.S. Supreme Court. The recordings were transcribed and manually word-aligned as part of the OYEZ\footnote{\url{http://www.oyez.org/}} project, then forced aligned and QCed at LDC.
        \item {\it SLX} \\
            SLX (LDC2003T15) is a corpus of sociolinguistic interviews conducted in the 1960s and 1970s by Bill Labov and his students. The interview subjects range in age from 15 to 81 and represent a diverse sampling of ethnicities, backgrounds, and dialects (e.g., southern Amercian English, African American English, northern England, and Scotland). While the recordings have good sound quality for field recordings (especially from that era), they were collected in a range of environments ranging from noisy homes (e.g., small children running around in the background) to public parks to gas stations.
        \item {\it VAST} \\
            The Video Annotation for Speech Technologies (VAST) corpus is a (mostly) unexposed collection of approximately 2,900 hours of web videos (e.g., YouTube and Vimeo) intended for development and evaluation of speech technologies; in particular, speech activity detection (SAD), diarization, language identification (LID), speaker identification (SID), and speech recognition (STT). Collection emphasized videos where people are talking with a particular emphasis on videos where the speakers spoke primarily English, Mandarin, and Arabic, which comprise the bulk of the corpus\footnote{Eight languages are represented in total: Arabic, English, Mandarin, Min Nan, Spanish, Portuguese, Russian, and Polish.}. Portions of this corpus have been exposed previously as part of the NIST 2017 Speech Analytic Technologies Evaluation, the  NIST 2017 Language Recognition Evaluation, NIST 2018 Speaker Recognition Evaluation, and DIHARD I and II.
        \item {\it YouthPoint} \\
            YouthPoint is an unpublished LDC corpus consisting of episodes of YouthPoint, a late 1970s radio program run by students at the University of Pennsylvania. The show had an interview format similar to shows such as NPR's Fresh Air and consisted of interviews between University of Pennsylvania students and various popular figures. The recordings were conducted in a studio on open reel tapes and later digitized and transcribed at LDC.
    \end{itemize}

\newpage
\section{Speech segmentation label files}
\label{app:sad}
For each recording, the reference speech segmentation will be provided via an HTK label file listing one segment per line, each line consisting of three space-delimited fields:
    \begin{itemize}
        \item segment onset in seconds from beginning of recording
        \item segment offset in seconds from beginning of recording
        \item segment label (always ``speech'')
    \end{itemize}
For example:
    \begin{itemize}
        \item[] 0.10 1.41 speech
        \item[] 1.98 3.44 speech
        \item[] 5.0 7.52 speech
    \end{itemize}
The segments in these files are guaranteed to be disjoint and to not extend beyond the boundaries of the recording session.

\newpage
\section{RTTM File Format Specification}
\label{app:rttm}
Systems should output their diarizations as Rich Transcription Time Marked (RTTM) files. RTTM files are text files containing one turn per line, each line containing ten space-delimited fields:

\begin{itemize}
    \item Type  --  segment type; should always by ``SPEAKER''
    \item File ID  --  file name; basename of the recording minus extension (e.g., ``rec1\_a'')
    \item Channel ID  --  channel (1-indexed) that turn is on; should always be ``1''
    \item Turn Onset  --  onset of turn in seconds from beginning of recording
    \item Turn Duration  -- duration of turn in seconds
    \item Orthography Field --  should always by ``$<$NA$>$''
    \item Speaker Type  --  should always be ``$<$NA$>$''
    \item Speaker Name  --  name of speaker of turn; should be unique within scope of each file
    \item Confidence Score  --  system confidence (probability) that information is correct; should always be ``$<$NA$>$''
    \item Signal Lookahead Time  --  should always be ``$<$NA$>$''
\end{itemize}

For instance:
\begin{itemize}
    \item []SPEAKER CMU_20020319-1400_d01_NONE 1 130.430000 2.350 $<$NA$>$ $<$NA$>$ juliet $<$NA$>$ $<$NA$>$
    \item[] SPEAKER CMU_20020319-1400_d01_NONE 1 157.610000 3.060 $<$NA$>$ $<$NA$>$ tbc $<$NA$>$ $<$NA$>$
    \item[] SPEAKER CMU_20020319-1400_d01_NONE 1 130.490000 0.450 $<$NA$>$ $<$NA$>$ chek $<$NA$>$ $<$NA$>$
\end{itemize}

\newpage
\section{UEM File Format Specification}
\label{app:uem}
Un-partitioned evaluation map (UEM) files are used to specify the scoring regions within each recording. For each scoring region, the UEM file contains a line with the following four space-delimited fields
\begin{itemize}
    \item File ID  --  file name; basename of the recording minus extension (e.g., ``rec1_a'')
    \item Channel ID  --  channel (1-indexed) that scoring region is on
    \item Onset  --  onset of scoring region in seconds from beginning of recording
    \item Offset  --  offset of scoring region in seconds from beginning of recording
\end{itemize}

For instance:
\begin{itemize}
    \item[] CMU_20020319-1400_d01_NONE 1 125.000000 727.090000
    \item[] CMU_20020320-1500_d01_NONE 1 111.700000 615.330000
    \item[] ICSI_20010208-1430_d05_NONE 1 97.440000 697.290000
\end{itemize}

\newpage
\section{Data Resources for Training}
\label{app:data}
This appendix identifies a (non-exhaustive) list of publicly available corpora suitable for system training.

\vspace{0.5cm}
\noindent{\bf Corpora containing meeting speech} \\
{\it LDC corpora}
\begin{itemize}
    \item ICSI Meeting Speech Speech (LDC2004S02)
    \item ICSI Meeting Transcripts (LDC2004T04)
    \item ISL Meeting Speech Part 1 (LDC2004S05)
    \item ISL Meeting Transcripts Part 1 (LDC2004T10)
    \item NIST Meeting Pilot Corpus Speech (LDC2004S09)
    \item NIST Meeting Pilot Corpus Transcripts and Metadata (LDC2004T13)
    \item 2004 Spring NIST Rich Transcription (RT-04S) Development Data (LDC2007S11)
    \item 2004 Spring NIST Rich Transcription (RT-04S) Evaluation Data (LDC2007S12)
    \item 2006 NIST Spoken Term Detection Development Set (LDC2011S02)
    \item 2006 NIST Spoken Term Detection Evaluation Set (LDC2011S03)
    \item 2005 Spring NIST Rich Transcription (RT-05S) Evaluation Set (LDC2011S06)
\end{itemize}

\vspace{0.25cm}
\noindent{\it Non-LDC corpora}
\begin{itemize}
    \item Augmented Multiparty Interaction (AMI) Meeting Corpus (\url{http://groups.inf.ed.ac.uk/ami/corpus/})
    \item CSTR VCTK Corpus (\url{https://homepages.inf.ed.ac.uk/jyamagis/page3/page58/page58.html})
\end{itemize}

\vspace{0.5cm}
\noindent{\bf Conversational telephone speech (CTS) corpora} \\
{\it LDC corpora}
\begin{itemize}
    \item CALLHOME Mandarin Chinese Speech (LDC96S34)
    \item CALLHOME Spanish Speech (LDC96S35)
    \item CALLHOME Japanese Speech (LDC96S37)
    \item CALLHOME Mandarin Chinese Transcripts (LDC96T16)
    \item CALLHOME Spanish Transcripts (LDC96T17)
    \item CALLHOME Japanese Transcripts (LDC96T18)
    \item CALLHOME American English Speech (LDC97S42)
    \item CALLHOME German Speech (LDC97S43)
    \item CALLHOME Egyptian Arabic Speech (LDC97S45)
    \item CALLHOME American English Transcripts (LDC97T14)
    \item CALLHOME German Transcripts (LDC97T15)
    \item CALLHOME Egyptian Arabic Transcripts (LDC97T19)
    \item CALLHOME Egyptian Arabic Speech Supplement (LDC2002S37)
    \item CALLHOME Egyptian Arabic Transcripts Supplement (LDC2002T38)
    \item Switchboard-1 Release 2 (LDC97S62)
    \item Fisher English Training Speech Part 1 Speech (LDC2004S13)
    \item Fisher English Training Speech Part 1 Transcripts (LDC2004T19)
    \item Arabic CTS Levantine Fisher Training Data Set 3, Speech (LDC2005S07)
    \item Fisher English Training Part 2, Speech (LDC2005S13)
    \item Arabic CTS Levantine Fisher Training Data Set 3, Transcripts (LDC2005T03)
    \item Fisher English Training Part 2, Transcripts (LDC2005T19)
    \item Fisher Levantine Arabic Conversational Telephone Speech (LDC2007S02)
    \item Fisher Levantine Arabic Conversational Telephone Speech, Transcripts (LDC2007T04)
    \item Fisher Spanish Speech (LDC2010S01)
    \item Fisher Spanish - Transcripts (LDC2010T04)
\end{itemize}

\vspace{0.5cm}
\noindent{\bf Other corpora} \\
{\it LDC corpora}
\begin{itemize}
    \item Speech in Noisy Environments (SPINE) Training Audio (LDC2000S87)
    \item Speech in Noisy Environments (SPINE) Evaluation Audio (LDC2000S96)
    \item Speech in Noisy Environments (SPINE) Training Transcripts (LDC2000T49)
    \item Speech in Noisy Environments (SPINE) Evaluation Transcripts (LDC2000T54)
    \item Speech in Noisy Environments (SPINE2) Part 1 Audio (LDC2001S04)
    \item Speech in Noisy Environments (SPINE2) Part 2 Audio (LDC2001S06)
    \item Speech in Noisy Environments (SPINE2) Part 3 Audio (LDC2001S08)
    \item Speech in Noisy Environments (SPINE2) Part 1 Transcripts (LDC2001T05)
    \item Speech in Noisy Environments (SPINE2) Part 2 Transcripts (LDC2001T07)
    \item Speech in Noisy Environments (SPINE2) Part 3 Transcripts (LDC2001T09)
	\item Santa Barbara Corpus of Spoken American English Part I (LDC2000S85)
    \item Santa Barbara Corpus of Spoken American English Part II (LDC2003S06)
    \item Santa Barbara Corpus of Spoken American English Part III (LDC2004S10)
    \item Santa Barbara Corpus of Spoken American English Part IV (LDC2005S25)
    \item HAVIC Pilot Transcription (LDC2016V01)
    \item Nautilus Speaker Characterization (LDC2018S17)
    \item SRI Speech-Based Collaborative Learning Corpus (LDC2019S01)
\end{itemize}

\vspace{0.25cm}
\noindent{\it Non-LDC corpora}
\begin{itemize}
    \item AVA ActiveSpeaker (\url{http://research.google.com/ava/})
    \item AVA Speech (\url{http://research.google.com/ava/})
    \item Common Voice (\url{https://voice.mozilla.org/en/datasets})
    \item LibriSpeech (\url{http://www.openslr.org/12/})
    \item Speakers in the Wild (SITW) (\url{http://www.speech.sri.com/projects/sitw/})
    \item VoxCeleb (\url{http://www.robots.ox.ac.uk/~vgg/data/voxceleb/})
    \item VoxCeleb 2 (\url{http://www.robots.ox.ac.uk/~vgg/data/voxceleb/vox2.html})
    \item VoxConverse (\url{http://www.robots.ox.ac.uk/~vgg/data/voxconverse/})
\end{itemize}

\newpage
\section{System descriptions}
\label{app:system}
In order to allow proper interpretation of the evaluation results, each submitted system must be accompanied by a system description. System descriptions are expected to be of sufficient detail for a fellow researcher to understand the approach and data/computational resources used to train and run the system. In order to make the preparation and format as consistent as possible, participants should use the IEEE Conference proceedings templates:
    \begin{quote}
        \url{https://www.ieee.org/conferences/publishing/templates.html}
    \end{quote}
    
with the following document structure:

\begin{itemize}
    \item Section 1: Authors
    \item Section 2: Abstract
    \item Section 3: Notable highlights
    \item Section 4: Data resources
    \item Section 5: Detailed description of algorithm
    \item Section 6: Results on the development set
    \item Section 7: Hardware requirements
\end{itemize}

\vspace{0.5cm}
\noindent{\bf Section 1: Authors} \\
Listing of people whose contributions you wish acknowledged. This section is optional.

\vspace{0.5cm}
\noindent{\bf Section 2: Abstract} \\
A short (a few sentences) high-level description of the system.

\vspace{0.5cm}
\noindent{\bf Section 3: Notable highlights} \\
A brief summary of what is different or any notable highlights. Examples of highlights could be differences among systems submitted; novel or unusual approaches, or approaches/features that led to a significant improvement in system performance.

\vspace{0.5cm}
\noindent{\bf Section 4: Data resources} \\
This section should describe the data used for training including both volumes and sources. For LDC or ELRA corpora, catalog ids should be supplied. For other publicly available corpora (e.g., AMI) a link should be provided. In cases where a non-publicly available corpus is used, it should be described in sufficient detail to get the gist of its composition. If the system is composed of multiple components and different components are trained using different resources, there should be an accompanying description of which resources were used for which components.

\vspace{0.5cm}
\noindent{\bf Section 5: Detailed description of algorithm} \\
Each component of the system should be described in sufficient detail that another researcher would be able to reimplement it. You may be brief or omit entirely description of components that are standard (i.e., no need to list the standard equations underlying an LSTM or GRU). If hyperparameter tuning was performed, there should be detailed description both of the tuning process and the final hyperparameters arrived at.

We suggest including subsections for each major phase in the system. Suggested subsections:
\begin{itemize}
    \item signal processing  --  e.g., signal enhancement, denoising, source separation
    \item acoustic features  --  e.g., MFCCs, PLPs, mel fiterbank, PNCCs, RASTA, pitch extraction
    \item speech activity detection details  --  relevant only for tracks 2 and 4
    \item segment representation  --  e.g., i-vectors, d-vectors
    \item speaker estimation  --  how number of speakers was estimated if such estimation was performed
    \item clustering method  --  e.g., k-means, agglomerative
    \item resegmentation details
\end{itemize}

\vspace{0.5cm}
\noindent{\bf Section 6: Results on the development set} \\
Teams must report performance of the submission systems (both primary and contrastive) on the DIHARD III core and full development sets. Both DER and JER should be reported as output by the the official scoring tool. Optionally, teams may report results of additional experiments (e.g., results from additional systems or on additional datasets). Teams are encouraged to quantify the contribution of their major system components that they believe resulted in significant performance gains, if any.

\vspace{0.5cm}
\noindent{\bf Section 7: Hardware requirements} \\
System developers should report the hardware requirements for both training and at test time:
\begin{itemize}
    \item Total number of CPU cores used
    \item Description of CPUs used (model, speed, number of cores)
    \item Total number of GPUs used
    \item Description of GPUs used (model, single precision TFLOPS, memory)
    \item Total number of TPUs used
    \item Generations of TPUs used (e.g., v2 vs v3)
    \item Total available RAM
    \item Used disk storage
    \item Machine learning frameworks used (e.g., PyTorch, Tensorflow, CNTK)
\end{itemize}

System execution times to process the entire development set must be reported.

\end{appendices}

\end{document}